# Gel-Based Morphological Design of Zirconium Metal-organic Frameworks

Bart Bueken,[1] Niels Van Velthoven,[1] Tom Willhammar,[2,3] Timothée Stassin,[1] Ivo Stassen,[1] David A. Keen,[4] Rob Ameloot,[1] Sara Bals,[2] Dirk De Vos,[1,*] and Thomas D. Bennett[5,*]

[1] Centre for Surface Chemistry and Catalysis, Department M²S, KU Leuven, Celestijnenlaan 200F p.o. box 2461, 3001 Leuven, Belgium

[2] EMAT, University of Antwerp, Groenenborgerlaan 171, 2020 Antwerp, Belgium

[3] Department of Materials and Environmental Chemistry, Berzelii Center EXSELENT on Porous Materials, Stockholm University, S-106 91 Stockholm, Sweden

[4] ISIS Facility, Rutherford Appleton Laboratory, Harwell Campus, Didcot, Oxon OX11 0QX, United Kingdom

[5] Department of Materials Science and Metallurgy, University of Cambridge, 27 Charles Babbage Road, Cambridge CB3 0FS, United Kingdom

## Abstract

The ability of metal-organic frameworks (MOFs) to gelate under specific synthetic conditions opens up new opportunities in the preparation and shaping of hierarchically porous MOF monoliths, which could be directly implemented for catalytic and adsorptive applications. In this work, we present the first examples of xero- or aerogel monoliths consisting solely of nanoparticles of several prototypical $Zr^{4+}$-based MOFs: UiO-66-X (X = H, $NH_2$, $NO_2$, $(OH)_2$), UiO-67, MOF-801, MOF-808 and NU-1000. High reactant and water concentrations during synthesis were observed to induce the formation of gels, which were converted to monolithic materials by drying in air or supercritical $CO_2$. Electron microscopy, combined with $N_2$ physisorption experiments, was used to show that an irregular nanoparticle packing leads to pure MOF monoliths with hierarchical pore systems, featuring both intraparticle micropores and interparticle mesopores. Finally, UiO-66 gels were shaped into monolithic spheres of 600 μm diameter using an oil-drop method, creating promising candidates for packed-bed catalytic or adsorptive applications, where hierarchical pore systems can greatly mitigate mass transfer limitations.

Metal-organic frameworks (MOFs) continue to attract significant academic interest by virtue of their large structural and chemical diversity.[1] These porous solids consist of inorganic metal-based nodes interconnected through coordinatively bonded organic linkers, and have been widely investigated as catalysts,[2,3] adsorbents,[4–7] drug delivery systems,[8,9] proton conductors[10] and sensing materials.[11,12] Accordingly, significant research has been oriented towards identifying and influencing the relationships between intrinsic MOF properties and targeted applications.

The crystallization of MOFs almost exclusively leads to polydisperse microcrystalline powders. While suitable for research purposes, this state limits the applicability of MOFs in industrial settings since the use of fine powders is associated with several technical challenges, including poor handling, dust formation, mass transfer limitations and strong pressure drops in packed beds.[13,14] To circumvent these issues, methods for the preparation of MOFs as meso- and/or macroscopically structured objects, preferably with hierarchical pore architectures, are highly sought after. Present approaches aimed at achieving this have mainly focused on structural templating, or on the formation of composite materials.[15–18]

Recently, zirconium-carboxylate MOFs, such as the Zr-terephthalate UiO-66 ($[Zr_6O_4(OH)_4(bdc)_6]$, UiO = Universitetet i Oslo; bdc = 1,4-benzenedicarboxylate) and its isoreticular derivatives, have risen to the forefront of the MOF field because of their excellent chemical and thermal stabilities, combined with high porosities and tunable properties.[19–21] Macroscale structuring of this subclass of MOFs has to date been achieved following two general strategies. First, Zr-MOFs have been deposited or grown onto support materials such as polymeric or ceramic monoliths, fibers and foams.[22–30] These composite materials combine the separation performance of the MOFs and the support's surface texture, but generally possess lower adsorption capacities due to the secondary component. A second approach involves pelletizing single component MOF powders via mechanical compression or extrusion. However, this avenue often results in pressure-induced losses of crystallinity and microporosity, and is unable to introduce meso- or macroporosity.[31–35] Binders are sometimes used to mitigate the first of these issues, however they diminish adsorption capacity, and judicious selection is needed to ensure their compatibility with the targeted application.[36–38]

A promising new route towards shaped, hierarchically porous, pure MOF materials starts with MOF gels and avoids the microcrystalline powder state altogether. Here, different from amorphous coordination polymer gels,[39,40] MOF nanoparticles crystallise during synthesis and aggregate to form a solid network throughout the synthesis solvent.[41–45] This results in a gel state of tunable viscosity, which adopts the shape of its container. Subsequent solvent removal results in nanoparticle agglomeration and the formation of monolithic xero- or aerogels. Li *et al.* for instance applied this strategy to produce a variety of monoliths constructed from Al-MOF nanoparticles.[42]

Sporadic examples of gel formation during the synthesis of UiO-66-type Zr-MOFs have been reported in the literature.[46-50] It has been documented how the synthesis of UiO-66-(COOH)$_2$ from ZrCl$_4$ in water proceeds through the formation of a white gel.[47,48] Gelation during the synthesis scale-up of UiO-66, using zirconyl chloride octahydrate (ZrOCl$_2$·8H$_2$O) as precursor, was also reported by Ragon and coworkers.[49] Finally, Liu *et al.* described a MOF gel synthesized from an ethanol-DMF mixture using 2-aminoterephthalic acid and ZrCl$_4$.[50] However, detailed investigations into the synthesis or structure of such gels are sparse; there are no reports on formation of crystalline monolithic materials or hierarchically porous architectures from these gels.

In this contribution, we uncover some of the key parameters controlling gelation of UiO-66, and transpose these across the Zr-MOF family to functionalized UiO-66 materials, UiO-67, MOF-801, MOF-808 and NU-1000. The resulting gels can be easily manipulated and solvent removal by drying in air or under supercritical $CO_2$ leads to the first reported hierarchically porous, monolithic Zr-MOF xero- and aerogels. As a proof-of-principle for the potential of these gels in an industrially relevant shaping process, monolithic UiO-66 spheres are prepared and characterized by oil-drop granulation of a UiO-66 gel.

## Results

### Gel formation

As a prototypical Zr-MOF system, we selected UiO-66 for our initial investigation on gel-based monolith formation, and performed a screening of several synthetic parameters to determine their influence on gelation (Table S1). For each synthesis, a fixed molar ratio of 1.45 $H_2$bdc:Zr ($H_2$bdc = 1,4-benzenedicarboxylic acid) was employed and the macroscopic outcome was visually evaluated after two hours of reaction at 100 °C. Three parameters were found to play a crucial role: (1) metal source, (2) reactant concentration and (3) the presence of water. First, the choice of the metal source strongly determined whether or not gelation occurred. Under comparable conditions, syntheses employing $ZrOCl_2 \cdot 8H_2O$ formed gels much more readily than those using $ZrCl_4$, which tended to yield microcrystalline precipitates (e.g. Table S1, entries 6, 8 vs. 21, 26 respectively). Note that, where needed, additional water and HCl were added to compensate for the 8 molar equivalents of crystallization water associated with $ZrOCl_2 \cdot 8H_2O$, and for the additional two equivalents of HCl produced from the hydrolysis of $ZrCl_4$. Secondly, increasing reactant concentrations (Table S1, entries 1 to 9) afforded progressively more 'non-flowing' gels. This effect was most pronounced when $ZrOCl_2 \cdot 8H_2O$ was employed as metal source. For instance, at a DMF:Zr ratio of ~1500 (8.7 mM $ZrOCl_2 \cdot 8H_2O$), a microcrystalline precipitate was obtained. Decreasing this ratio to 620 and 388 led to viscous solutions with a gel-like consistency, but which flowed downward upon turning the synthesis vessel upside down. Starting from DMF:Zr ratios below 200, 'non-flowing' gels were obtained (Figure 1, a). For syntheses based on $ZrCl_4$, an increase in reactant concentration alone was insufficient to induce gelation (entries 17, 20, 26 and 30). Rather, a combination between high reactant concentrations and addition of water was decisive for these reaction mixtures, with an increase in the latter clearly steering the syntheses towards more 'non-flowing' gels (Table S1, entries 19-23, 24-25, 26-27, 28-31). However, the concentration of water seemed to have a far less pronounced effect on syntheses starting from $ZrOCl_2 \cdot 8H_2O$. Curiously, acetic acid, often used as a synthesis modulator to facilitate Zr-MOF crystallization,[51] did not appear to have any noticeable macroscopic effect on the obtained gel products at the employed acetic acid:Zr ratio of 3.5 (Table S1, entries 4, 9, 18 vs. 5, 10, 19, respectively).

Following synthesis, X-ray diffraction patterns were recorded for each of the gels to confirm the formation of UiO-66 (Figure 2). The diffraction pattern presented in Figure 2, c, which is representative for all formed gels, contains two Bragg reflections centred around 7 ° and 8.5 ° 2θ, which correspond to the (111) and (200) reflections of the UiO-66 structure.[19] Furthermore, their broadness indicates the presence of UiO-66 nanoparticles, with domain sizes between 10 to 15 nm as determined by the Scherrer equation. As expected, syntheses that did not yield gels produced microcrystalline UiO-66 as a powder precipitate (Figure 2, b).

**Monolith formation and characterization**

We subsequently directed our efforts to transforming the UiO-66 gels into dry, monolithic solids. Prior to solvent evacuation, unreacted linkers were removed from the gel matrices through solvent exchange, by dispersing an as-synthesized gel in an equal volume of fresh DMF, followed by a shear-induced homogenization using a vortex mixer. This readily turned the 'non-flowing' system into a volume-expanded, 'flowing' gel of lower viscosity, which was allowed to rest overnight. Subsequently, this system could again be converted to a 'non-flowing', compacted state through a centrifugation-driven syneresis. The excess solvent was phase-separated from the gel during centrifugation, and could easily be decanted. This whole process was repeated several times using fresh solvent (DMF or ethanol), until after the final step, an ethanol-containing, 'non-flowing' state was acquired, with a volume adjusted to be equal to the volume of the as-synthesized gel.

Monolithic, optically transparent xerogels were obtained by drying ethanol-exchanged gels in air at 200 °C (Figure 1, b, c; Figure S1). Alternatively, solvent extraction using supercritical $CO_2$ afforded aerogel monoliths (Figure S2). For the latter, no change in volume was observed upon solvent removal, whereas the xerogels underwent significant shrinkage due to the capillary forces exerted during the drying process. Regardless of the drying method, the broadened X-ray diffraction pattern of UiO-66 present in the precursor gels was retained in each case (Figure 2, d-e). While X-ray diffraction only gives information on the average, long-range structure of the monolith, additional information on atomic length scales was obtained by extracting the atomic pair distribution function (PDF) from X-ray total scattering experiments. A PDF provides a distribution of pairwise interatomic distances within a sample, and as such gives more insight into its local structure. Figure S3 displays the PDFs of a representative air-dried xerogel and a microcrystalline UiO-66 powder recorded under the same conditions. The main peaks, present in both PDFs at interatomic distances of 2.33 Å, 3.75 Å and 4.89 Å, correspond to UiO-66's intracluster Zr-O and Zr-Zr atom pairs, while those at greater interatomic distances can be attributed to correlations between atoms in neighbouring clusters.[34,52,53] For the peaks observed below interatomic distances of 1 Å, the quality of lab-recorded total scattering data is typically insufficient to allow for a reliable interpretation.[54] The good agreement between both PDFs, as well as with previously reported PDFs for UiO-66,[34,52,53] further confirms the successful crystallization of UiO-66 during gel formation. However, while there are no direct indications that the gels contain additional phases, their presence could not be ruled out entirely based on scattering data alone (both PDF and diffraction data), especially since the nearest-neighbour Zr-Zr and Zr-O distances in $ZrO_2$ are almost identical to those in UiO-66.[55]

Thus, to gain more insight into the micro/meso-structure and phase-purity of the MOF gels, a typical air-dried xerogel sample (formed using the gel in Table S1, entry 10) was selected and investigated using electron microscopy (Figure 3). The annular dark field (ADF) scanning transmission electron microscopy (STEM) micrographs of this sample are shown in Figure 3, b-e, and reveal that the monoliths are made up entirely of aggregated nanoparticles of approximately 10 nm in size. Both selected area electron diffraction (SAED) and Fourier transformations from individual nanocrystallites in the ADF-STEM images (Figure 3, d-f) show these particles to have diffraction features with *d*-values of 11.5-12.1 Å and 10.5 Å. These coincide with the (111) and (200) reflections of the face-centred cubic UiO-66 crystal structure,[19] essentially confirming the observations made from powder X-ray diffraction and PDF analysis. Furthermore, no additional phases, such as amorphous or crystalline zirconium oxides, were observed in these experiments.

From the images in Figure 3 (a-d), it is clear that the randomly packed crystallites generate mesoporous interparticle voids throughout the monolithic structure. An electron tomographic reconstruction enabled the 3-dimensional (3D) visualization of these voids in a single aggregate xerogel particle of ~100 nm wide (Figure 4 a, b), illustrating how mesopores of 10-30 nm diameter permeate throughout the entire particle. Macroscopically, the porosity of the monoliths was confirmed by $N_2$ physisorption experiments (Figure 4 c, Table 1). For both the xero- (red) and aerogel (blue), the isotherms feature a steep increase in $N_2$ uptake at low $p/p^0$, signifying adsorption in the micropores of the UiO-66 framework; at high $p/p^0$, the behaviour is reminiscent of a IUPAC type IV isotherm, as indicated by the hysteretic desorption isotherm characteristic of mesoporous materials.[56] For the air-dried UiO-66 xerogel considered above, a total pore volume of 2.09 $cm^3 \cdot g^{-1}$ was measured, with a calculated Brunauer-Emmett-Teller (BET) surface area of 1459 $m^2 \cdot g^{-1}$ and a micropore area of 844 $m^2 \cdot g^{-1}$, as determined by the t-plot method. Applying the Barrett-Joyner-Halenda (BJH) method revealed fairly uniform mesopores of approximately 15 nm in diameter in the xerogel (Figure S4a), which matches the range observed by electron microscopy. The micropore size distribution (Figure S4b) further shows the presence of 6 Å and 8 Å diameter pores, consistent with the cages present in the UiO-66 structure.[19] Similar results were obtained for the aerogel, with a total pore volume of 1.66 $cm^3 \cdot g^{-1}$, and BET and micropore surface areas of 1255 $m^2 \cdot g^{-1}$ and 847 $m^2 \cdot g^{-1}$, respectively. We ascribe this slightly lower total pore volume to incomplete pore evacuation of the aerogel, as indicated by the thermogravimetric analysis (TGA) traces for the samples (Figure S5). The BJH pore size distribution however is much broader than for the xerogel, and is centred on 24 nm, consistent with the absence of capillary forces during supercritical $CO_2$ drying. Overall, the porosity of the UiO-66 monoliths far exceeds that of bulk UiO-66 powder by a factor of 3-4 (Table 1). The somewhat lower micropore surface area of the gels compared to that of a typical bulk UiO-66 powder (1085 $m^2 \cdot g^{-1}$) likely finds its origin in the crystallite size; as size decreases, micropore surface area and volume is lost relative to the external surface area.[57,58] These values are in agreement with the results obtained by Taddei *et al.*,[59] who reported micropore areas of 966 $m^2 \cdot g^{-1}$ and 730 $m^2 \cdot g^{-1}$ for UiO-66 nanoparticles of 25 and 10 nm, respectively.

Both xero- and aerogels exhibit excellent thermal stability based on thermogravimetric analysis (TGA). As illustrated in Figure S5, both materials feature a single decomposition step starting at ~450 °C ($O_2$ atmosphere), corresponding to the disintegration of the framework's bdc linkers. From these data, an average of approximately 11 linkers per $Zr_6$-cluster was found for the xerogel, which is in line with the excess of $H_2$bdc employed to synthesize the gels.[60] For the aerogel, an average of 11.8 linkers per cluster, and an additional mass loss step between 100 °C and 200 °C was observed, suggesting the presence of residual DMF and/or $H_2$bdc not removed by the post-treatment. While maintaining the same thermal stability as bulk UiO-66,[19,60] the hierarchical nature of the gels results in a lower mechanical stability relative to 'pristine' UiO-66. For the air-dried xerogel sample, an elastic modulus ($E$) between 9.3 (±0.3) GPa and 10.5 (±0.5) GPa was found by nanoindentation experiments on polished monolithic particles (Figure S6-S7). While difficult to compare with literature data due to the limited number of reports and the defect-dependence of UiO-66's mechanical properties, at least one study indicates this to correspond to about 20% - 33% of the value determined for UiO-66 with a similar number of missing linkers.[61]

**Mesoporous monolithic spheres**

To illustrate the potential of Zr-MOF gels to be shaped into a variety of monolithic objects, we aimed to prepare spherical, monodisperse monoliths of UiO-66 based on the industrially employed oil-drop granulation process,[62] which prepares spherical silica or alumina granules by dripping sol droplets into an immiscible hot oil, followed by *in situ* gelation. A similar oil-drop setup was constructed in-house (see Supplementary Information) to dispense a UiO-66 gel. While it was equally possible to perform the oil-drop shaping directly from the gel's synthesis solution, as in the conventional process, we found that using a preformed gel allowed for a more controlled shaping process and significantly simplified the washing procedures following bead formation, since the starting gel was already extensively solvent exchanged with DMF to remove excess reactants.

A regular UiO-66 gel (Table S1, entry 10) was synthesized, and solvent exchanged five times with fresh DMF to remove any unreacted linkers. After the final washing iteration, the volume of the gel was again expanded, in this case with 30 wt% of fresh DMF, yielding a mildly 'flowing' gel of which droplets could be dispensed by a perfusor pump into a flow of an immiscible, heated silicone oil. The spherical gel droplets subsequently underwent a temperature-induced syneresis to a xerogel over the course of 10 minutes at 150 °C. After collection and separation from the oil phase, the monolithic spheres were washed several times with dichloromethane to remove any oil residues, followed by a final annealing step (200 °C, air, 1 h). The resulting spheres (Figure 5, a, b) were uniform in size, which could be varied by changing the diameter of the dispensing needle. For instance, xerogel spheres with ~600 μm in diameter were obtained by using a 1.2 mm diameter dispensing needle, which produced gel droplets of 2 mm in diameter. Thus, the droplets underwent a ~37-fold volume shrinkage while transitioning to the xerogel spheres. X-ray diffraction (Figure 5, e) confirmed that the UiO-66 nanoparticles maintain their structure. The internal architecture of the spheres was investigated by scanning electron microscopy (Figure 5, c, d). Similar to the xerogel samples prepared by drying the UiO-66 gel in air, interparticle mesopores could be seen on the surface and in cross-sections of the spheres due to an irregular nanoparticle packing. The micro/mesoporous $N_2$ physisorption isotherm (Figure 5, f; Table 1) corroborated the presence of mesopores, narrowly distributed around a diameter of 14.7 nm, and a total pore volume and BET surface area of respectively 1.9 cm$^3$·g$^{-1}$ and 1127 m$^2$·g$^{-1}$ were found, essentially matching those found for the air-dried xerogel.

## Discussion

The observation that high concentrations of water, linker and metal source in the synthesis of UiO-66 induce the formation of a nanocrystalline gel state leads us to suggest a rapid and excessive crystal nucleation to be at the origin of gelation. Indeed, this hypothesis can be rooted in classical crystallization theory, which models the nucleation rate to be exponentially dependent on the reactant supersaturation. Furthermore, a prerequisite for the UiO-66 framework to form is the formation of its $[Zr_6O_4(OH)_4]^{12+}$ clusters through hydrolysis of the employed Zr-salt.[51] Hence, the concentration of water greatly influences the crystallization of UiO-66. For instance, Schaate et al. reported how increasing the amount of water present in (diluted) synthesis media yielded progressively smaller UiO-66 crystallites, with sizes down to 14 nm.[51] Similarly, Ragon et al. found UiO-66 to crystallise significantly faster in the presence of water and attributed this to an easier formation of the $Zr_6$-clusters.[49] The observed differences in gelation between $ZrCl_4$ and $ZrOCl_2·8H_2O$ can be interpreted in a similar fashion, since $ZrOCl_2·8H_2O$ already is the primary hydrolysis product of $ZrCl_4$ and occurs as the tetranuclear cluster $[Zr_4(OH)_8(H_2O)_{16}]Cl_8·12(H_2O)$; the latter can be considered a direct precursor for the eventual $Zr_6$-clusters in UiO-66.[63] Thus, syntheses utilizing $ZrOCl_2·8H_2O$ are less sensitive to the addition of extra water due to its more advanced hydrolysis degree.[49]

We propose gelation to be a direct consequence of the high concentration of formed UiO-66 nanoparticles, which aggregate primarily through non-covalent Van der Waals interactions, although some degree of coordinative cross-linking or intergrowth between crystallites cannot be ruled out. The resulting colloidal suspension contains a weakly bound network of solids throughout its entire volume, and effectively adopts a gel-like state (Figure S8). Because of a rapid decrease in reactant concentration, and concomitant increase in solution viscosity, further growth of the UiO-66 crystallites is likely impeded, leading to a kinetically stable state. In more dilute systems, the smaller number of nuclei rather continues to grow and precipitates as a microcrystalline MOF powder. While similar gelation mechanisms have been proposed by others for MOF gels based on di- and trivalent cations,[41,42,44,45] it should be noted that in several of these gels, additional, non-crystalline phases act as a binder between the MOF nanoparticles,[42] or as a scaffold in which they are embedded.[44] In case of the UiO-66 gels presented here, the available evidence points to the absence of such phases (Figure 3).

Associated with high particle concentrations in the UiO-66 gel is a viscoelastic behaviour, with viscous flow occurring only above a certain yield stress at which sufficient interparticle interactions are overcome. The observed differences between 'non-flowing' and 'flowing' gels find their origin in a lower viscosity and yield stress for the latter, likely resulting from a lower volume fraction of crystallites and/or larger crystallite sizes, allowing the gel to flow under gravitational forces.[13] Post-synthetically manipulating the solid's concentration thus offers a straightforward means to controlling the viscosity and state of the system: applying shear forces, as in vortex mixing, enables the solid network of a 'non-flowing' gel to be broken and redispersed in a larger solvent volume; the concentration of particles is lowered, and the viscoelastic properties of the system resemble that of a 'flowing' gel. Conversely, centrifugation of a 'flowing' gel, followed by removal of excess solvent achieves the opposite transformation (Figure S8), and allowed us to obtain 'non-flowing' gels from as-synthesized 'flowing' gels.

Since the inorganic $Zr_6$-cluster is shared by many Zr-MOFs, we hypothesized that a rational choice of synthetic conditions, followed by application of the gelation principles outlined here for UiO-66, would allow the extrapolation of gel formation to other $Zr_6$-cluster systems. Starting from the routine used to synthesize the UiO-66 monoliths (Table S1, entry 10), 'non-flowing' gels of the isoreticular MOFs UiO-66-$NO_2$, UiO-66-$NH_2$, UiO-66-$(OH)_2$, UiO-67 and MOF-801[64] were prepared by simply substituting $H_2$bdc for equimolar amounts of their respective linkers (Figure S9). MOF-808[64] and NU-1000[65] are two Zr-MOFs based on respectively 1,3,5-benzenetricarboxylate and 1,3,6,8-tetrakis(*p*-benzoate)pyrene as linkers and feature topologies that significantly deviate from the face-centered cubic UiO-66 structure. Gels of these MOFs were still readily prepared by both increasing the concentration of reactants and including additional water relative to their original syntheses. In each case, an X-ray diffraction pattern (Figure S10) matching that of the desired phase, but with broad reflections indicative of nanosized crystallites, was obtained. Furthermore, each of these gels could be solvent exchanged following the procedure established for UiO-66, and transformed into monolithic xerogels by drying in air (Figure S9).

In conclusion, a new method to structure UiO-66 at the mesoscale is presented, by steering the synthesis towards a MOF-nanoparticle based gel state. The UiO-66 gels could be transformed to hierarchically porous monoliths, both as xero- and aerogels, with pore volumes (2.09 $cm^3 \cdot g^{-1}$ and 1.66 $cm^3 \cdot g^{-1}$, respectively) far exceeding those of bulk UiO-66 powder. Furthermore, the UiO-66 gel state can be exploited to form shaped objects, as exemplified by the mesoporous, binder-free UiO-66 spheres prepared here by an industrially relevant oil-drop granulation. Since gelation is achieved in conditions which enhance the formation rate of the ubiquitous $Zr_6$-clusters, the principles outlined for preparing UiO-66 gels can be extrapolated to form a variety of other $Zr_6$-cluster based MOF gels, as shown for isoreticular analogues of UiO-66, MOF-808 and NU-1000. Hence, the gels and monoliths presented here provide a step towards shaping Zr-MOF materials for applications in catalysis or adsorption. Finally, the optically transparent nature of the xerogels may be of interest in the preparation of transparent films and coatings.

## Methods

**Gel and monolith synthesis**

A typical UiO-66 gel synthesis (e.g. Table S1, entry 10), was performed in a 100 mL pyrex Schott bottle by dissolving 14.5 mmol (2.41 g) $H_2$bdc (98%, Sigma Aldrich) and 10 mmol (3.22 g) $ZrOCl_2·8H_2O$ (>98%, Acros) in 60 mL DMF (>99%, Acros), after which 1.5 mL of a 37 wt% HCl solution (Fisher) and 2 mL glacial acetic acid (Fisher) were added. The resulting solution was placed in a conventional synthesis oven at 100 °C for 2 hours. Similar gels could be obtained by adjusting the molar ratios of the employed reactants in the above-described procedure. UiO-66-$NO_2$, UiO-66-$NH_2$, UiO-66-$(OH)_2$, UiO-67 and MOF-801 gels were prepared by replacing $H_2$bdc in the procedure outlined above with an equimolar amount of 2-nitroterephthalic acid, 2-aminoterephthalic acid, 2,5-dihydroxyterephthalic acid, 4,4'-biphenyldicarboxylic acid and fumaric acid, respectively. MOF-808 gels were obtained by dissolving 4.8 mmol (1.02 g) 1,3,5-benzenetricarboxylic acid (95%, Sigma Aldrich) and 3.3 mmol (1.07 g) $ZrOCl_2·8H_2O$ in a mixture of 10 mL DMF, 10 mL formic acid (98%, Fisher) and 0.3 mL distilled water. Gelation was induced by reacting this mixture at 100 °C in a conventional synthesis oven for 2 h. An NU-1000 gel was prepared by dissolving 167 µmol (53.7 mg) $ZrOCl_2·8H_2O$ in 1 mL of DMF. Following complete dissolution, 146 µmol (100 mg) of 1,3,6,8-tetrakis(*p*-benzoic acid)pyrene, synthesized as reported previously,[65] 25 µl of HCl (37 wt%) and 584 µmol (71.2 mg) of benzoic acid were added, after which the mixture was placed in an ultrasound bath for 15 min. Subsequently, a gel was obtained by reacting this mixture at 100 °C for 48 h. Microcrystalline UiO-66 powder was prepared by dissolving 1.172 g $ZrCl_4$ and 1.263 g $H_2$bdc in 150 mL of DMF. To this synthesis 0.75 mL HCl (37 wt%) and 1.5 mL glacial acetic acid were added, after which it was allowed to react in a conventional synthesis oven at 120 °C for 96 h. The powder product was recovered through centrifugation.

After synthesis, the obtained gels were washed twice with DMF and thrice with ethanol. In each step, fresh solvent was added so that the total gel volume was expanded to double that of the as-synthesized gel. Using a vortex mixer, the gels were homogenized with the fresh solvent, after which the expanded gels were allowed to rest overnight at 120 °C for DMF-exchanged gels, and 60 °C for ethanol-exchanged gels. Subsequently, the gels were centrifuged, after which the supernatant solution was decanted. Following the final washing step, the volume of the gel was adjusted to again by addition of fresh solvent, to achieve a volume equal to that of the as-synthesized gel.

To form monolithic xerogels, the ethanol-exchanged gels were placed in a glass petri dish or porcelain crucible in a synthesis oven at 200 °C for 2 hours, which resulted in the formation of transparent chunks of various morphologies. Aerogel monoliths were prepared by fully loading the sample chamber (5.3 cm$^3$) of a SCLEAD-2BD autoclave (KISCO) with an ethanol-exchanged gel, followed by supercritical $CO_2$ extraction at 14 MPa and 50 °C for 1 h and a final two-stage evacuation step at respectively 100 °C and 125 °C under 0.1 mbar for 6 h each.

The UiO-66 gel used for preparing the monolithic spheres was obtained as described above. Following five DMF washing steps, the gel was centrifuged once more, and redispersed with 30 wt% of fresh DMF after decanting the supernatant. A 3 mL syringe (BD Luer Lock) was filled with the final gel and placed in a perfusor pump (Braun B), attached to an in-house built flow setup (details provided in Supporting Information). After recovery from the collection vessel, the xerogel spheres were soaked several times in dichloromethane to remove excess silicone oil from their outer surface and pores, followed by a final annealing step at 200 °C in air 1 h.

**Material characterization**

X-ray diffraction patterns were recorded on a STOE COMBI P diffractometer (monochromated Cu K$_{\alpha1}$-radiation, $\lambda$ = 1.54060 Å) equipped with an IP-PSD detector in Bragg-Brentano transmission geometry. A PANalytical Ag-source X'pert Pro MPD lab diffractometer ($\lambda$ = 0.56089 Å) was used to collect room temperature X-ray total scattering data. Samples were loaded into 1.0 mm diameter quartz capillaries. The reciprocal space data within the range 0.7 < Q <~15 Å$^{-1}$ were corrected for background, multiple scattering, container scattering, and absorption using GudrunX, and the corresponding PDFs gained by Fourier transform.[66–68] Thermogravimetric analyses were performed on a TA instruments TGA Q500. Samples were heated at a rate of 5 °C·min$^{-1}$ to 650 °C under an $O_2$ flow.

Optical microscopy images were collected on a Leica DM 2000 microscope (Leica Microsystems) at 10x and 40x objective magnifications. Scanning electron microscopy images were recorded on a Jeol JSM-6010LV. The samples were sputtered with gold or palladium before loading into the microscope. ADF-STEM images and SAED patterns were collected using an FEI Tecnai transmission electron microscope operated at an acceleration voltage of 200 kV. The high resolution ADF-STEM images were acquired using an aberration corrected cubed FEI Titan microscope operated at 300 kV. The samples were prepared by crushing the monolith sample in ethanol and depositing drops of the suspension on a copper grid covered with a holey carbon film. The samples were additionally visualized using electron tomography. In order to handle the beam sensitive nature of the MOF monoliths, special care was taken during data acquisition. The tomographic reconstruction was performed based on a tilt series of 2D low-dose TEM images acquired between -70° and +60° with an increment of 5°. The tomographic reconstructions were performed using a total variation minimization algorithm,[69] and the visualization was done with the AMIRA software package.

$N_2$ physisorption measurements were performed on a Micromeritics 3Flex surface analyzer at liquid nitrogen temperature (-196 °C). Prior to the measurements, the samples (50-100 mg) were outgassed for 8 h at 125 °C and 0.1 mbar vacuum. Surface areas were calculated using the multi-point BET (Brunauer-Emmett-Teller) method applied to the isotherm adsorption branch, in line with the Rouquerol consistency criteria.[57] External surface areas and micropore volumes were calculated using the t-plot method (Harkins and Jura thickness equation; thickness range 3.5-5 Å). Micropore areas were obtained by subtracting the external surface area from the BET surface area. Barrett-Joyner-Halenda (BJH) pore size distributions were determined to characterize the mesopores, while the micropores were defined by the Tarazona Non-Local Density Functional Theory (NLDFT) pore size distribution model.

Nanoindentation experiments were performed using an MTS Nanoindenter XP, located in an isolation cabinet to shield against thermal fluctuations and acoustic interference. Before indentation, monolith surfaces were first cold-mounted using an epoxy resin and then carefully polished using increasingly fine diamond suspensions. Indentations were conducted under the dynamic displacement-controlled "continuous stiffness measurement" mode. The elastic modulus ($E$) was subsequently determined as a function of the surface penetration depth. A 2 nm sinusoidal displacement at 45 Hz was superimposed onto the system's primary loading signal, and the loading and unloading strain rates were set at $5 \times 10^{-2}$ s$^{-1}$. All tests were performed to a maximum indentation depth of 500 - 1000 nm using a Berkovich (i.e. three-sided pyramidal) diamond tip of radius ~100 nm. The raw data (load-displacement curves) obtained were analysed using the Oliver and Pharr method (Poisson's ratio set at 0.18).[70] Data resulting from surface penetrations of less than 100 nm were discarded due to imperfect tip-surface contacts.

## Acknowledgements

B. B., T.S. and I.S. acknowledge the FWO Flanders (doctoral and post-doctoral grants). T. W. acknowledges a post-doctoral grant from the Swedish Research Council. T. D. B. acknowledges the Royal Society (University Research Fellowship) and Trinity Hall (University of Cambridge) for funding. S.B. and D.D.V. are grateful for funding by Belspo (IAP 7/05 P6/27) and by the FWO Flanders. D.D.V further acknowledges funding from the European Research Council (project H-CCAT). S.B. acknowledges financial support from the European Research Council (ERC Starting Grant #335078-COLOURATOMS). The authors acknowledge Arnau Carne and Shuhei Furukawa for assistance with supercritical $CO_2$ extraction, and Charles Ghesquiere for assistance in synthesis.

## Author Contributions

**Competing Financial Interests**

The authors declare no competing financial interest.

## Materials & Correspondence

**Figures**

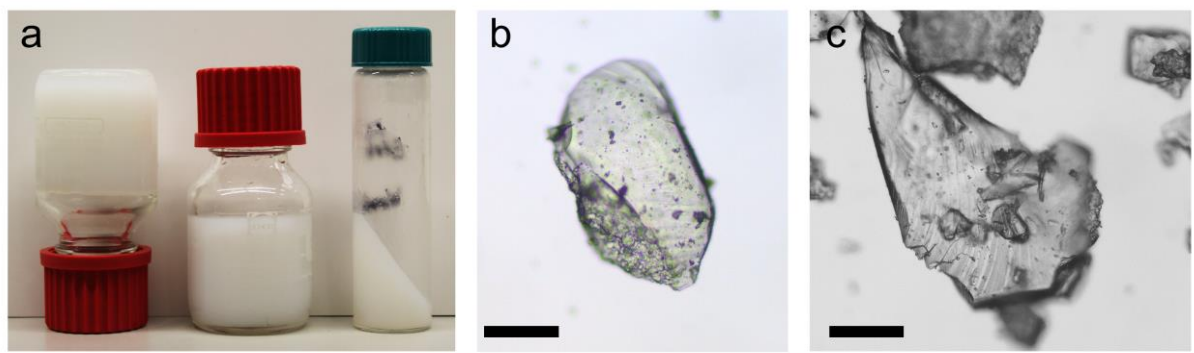

**Figure 1 | Gels and monolithic particles of UiO-66.** (**a**) 'Non-flowing' gels of UiO-66 synthesized from $ZrOCl_2 \cdot 8H_2O$ and $H_2bdc$ (Table S1, entry 10). (**b**, **c**) Optically transparent monolithic xerogel particles obtained by crushing 'non-flowing' gels dried in air at 200 °C. Prior to drying, the gels were washed and solvent-exchanged with ethanol. Scale bar = 100 μm (see also Figure S1).

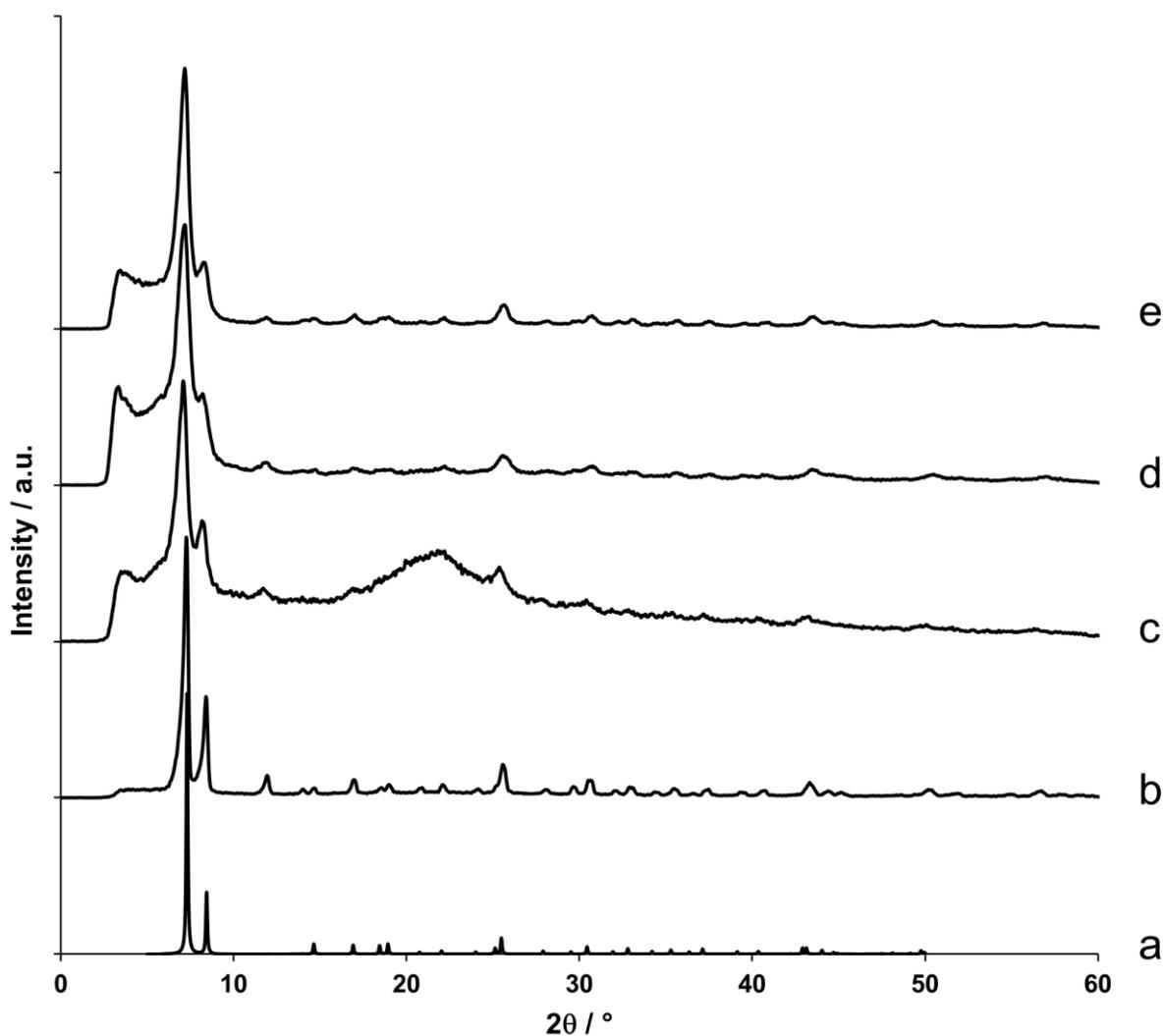

**Figure 2 | X-ray diffraction patterns of UiO-66 gels and monoliths.** (**a**) Simulated diffraction pattern of UiO-66.[19] (**b**) UiO-66 prepared as microcrystalline powder. (**c**) UiO-66 prepared as ethanol-exchanged gel (Table S1, entry 10). (**d**) Air-dried xerogel monolith prepared from the gel in (c). (**e**) Aerogel monolith obtained after supercritical $CO_2$ extraction of the gel in (c).

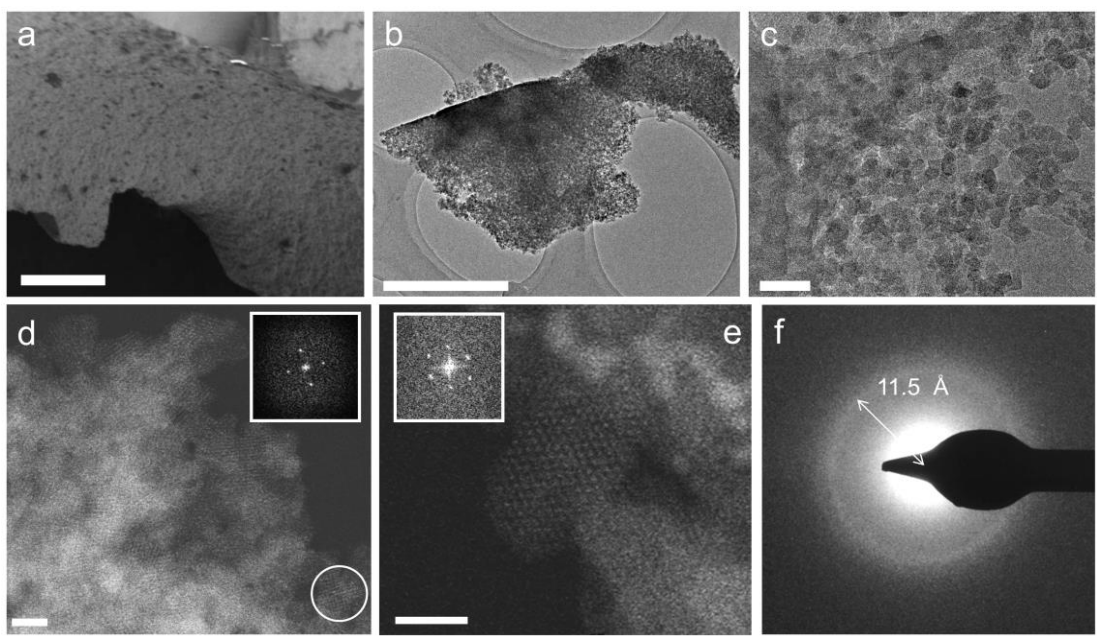

**Figure 3 | Electron Microscopy of monolithic UiO-66 xerogels.** (**a**) Scanning electron microscopy (SEM) image (scale bar = 25 µm) of xerogel particle. (**b**, **c**) TEM images (b, scale bar = 1 µm; c, scale bar = 50 nm) of xerogel particles, which consist of irregularly packed MOF nanoparticles with interparticle voids. (**d**) ADF-STEM image of the xerogel in (b), illustrating the crystalline nature of the nanoparticles. Bright contrast corresponds to areas of high density (scale bar = 10 nm). The inset shows the Fourier transform of the nanoparticle circled in white, consistent with the face-centered cubic lattice of UiO-66 viewed along the [100] direction. (**e**) ADF-STEM image of the sample in (c), showing an individual UiO-66 nanoparticle oriented along [110]. Each bright spot corresponds to a single column of [$Zr_6O_4(OH)_4(R-COO)_{12}$] clusters (scale bar = 10 nm). The Fourier transform of this nanoparticle (inset) features reflections that can be indexed as the (111) and (200) reflections of UiO-66, corresponding to *d*-spacings of 12.1 Å and 10.5 Å, respectively. (**f**) SAED pattern of the aggregate particle in (b). The diffraction ring corresponds to a *d*-spacing of ~11.5 Å, consistent with UiO-66's (111) reflection (12.0 Å).

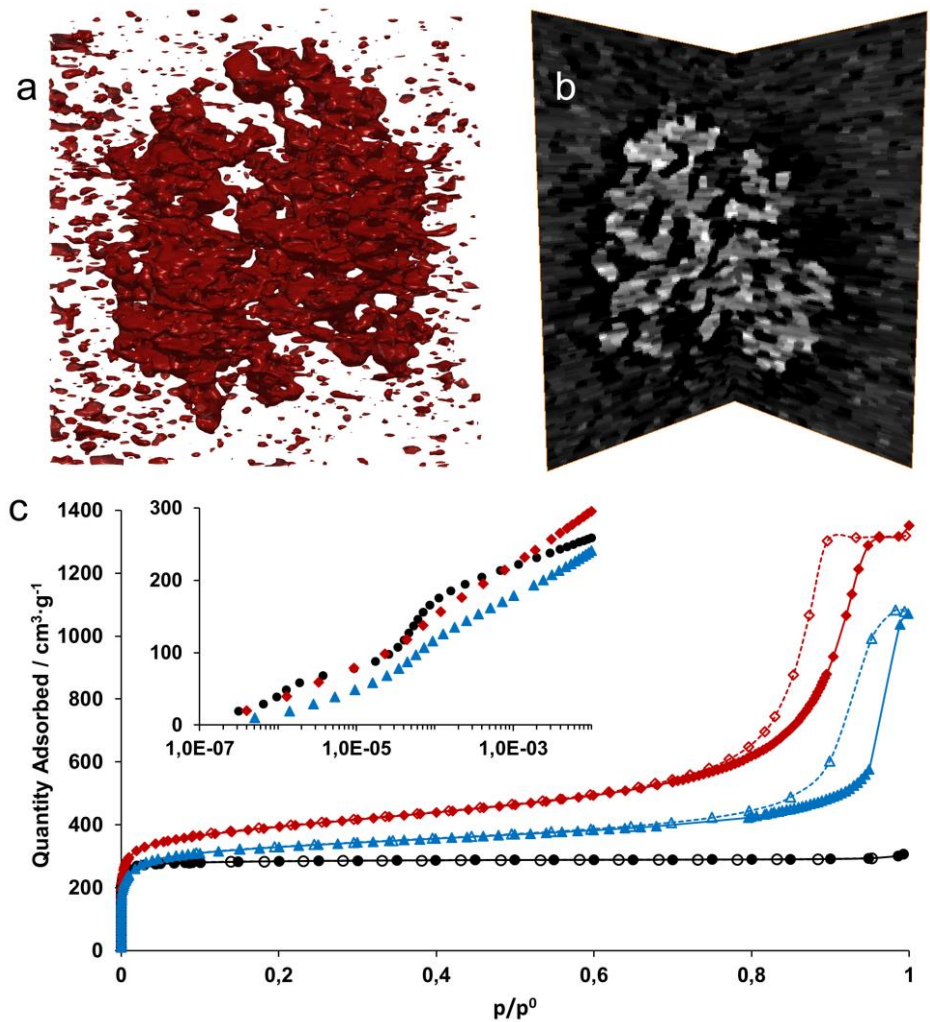

**Figure 4 | Hierarchical porosity in UiO-66 monoliths.** (**a**) Electron tomographic reconstruction of a single mesoporous monolithic xerogel particle (approx. 100 nm wide). Matter is represented in red. (**b**) Slice through the 3D reconstruction in (a). Bright contrast corresponds to matter, revealing intraparticle mesoporous voids. (**c**) Nitrogen physisorption isotherms (77 K) for a microcrystalline UiO-66 powder (black circles) and xerogel (red diamonds) and aerogel (blue triangles) monoliths (full symbols = adsorption branch; open symbols = desorption branch). The hysteretic desorption above $p/p^0 = 0.8$ is attributed to capillary condensation in the mesopores. The inset shows a logarithmic representation of the adsorption branch at low $p/p^0$. The two-step profile corresponds to the uptake of $N_2$ in the smaller tetrahedral (6 Å) and larger octahedral cages (8 Å), respectively.

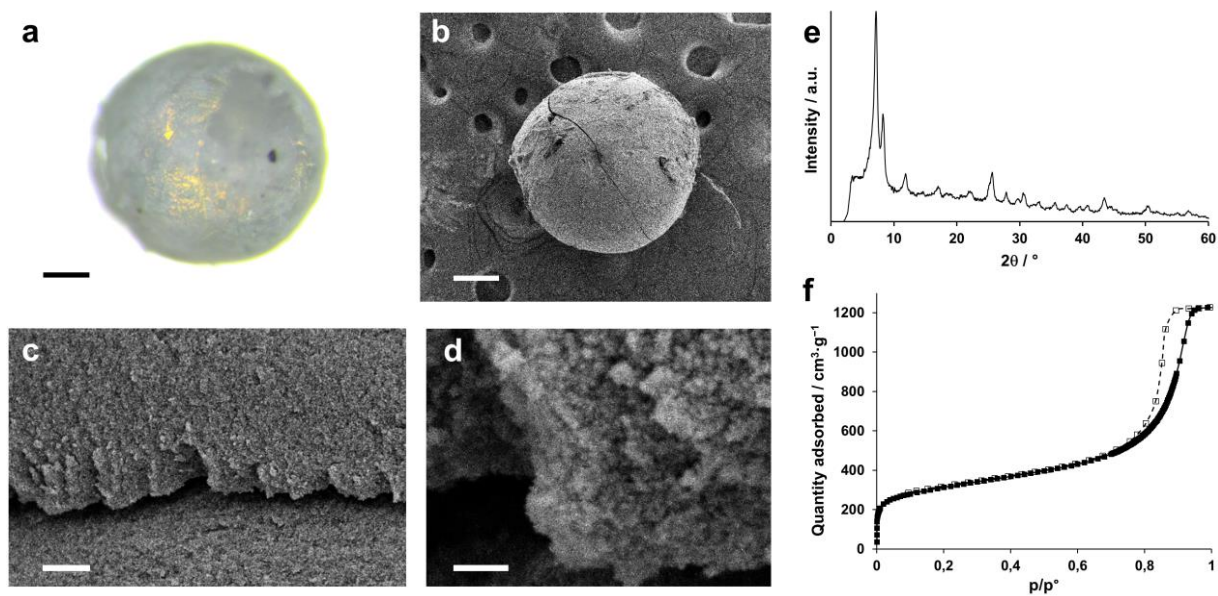

**Figure 5 | Monolithic UiO-66 xerogel spheres.** (**a**) Optical image of a single sphere (scale bar = 150 μm). (**b**) SEM micrograph of a single sphere (scale bar = 150 μm). (**c**) SEM micrographs of a cross-section of the interior sphere architecture (scale bar = 3 μm). (**d**) Close-up of (c), highlighting the nanoparticulate structure and mesoporosity (scale-bar = 0.5 μm). (**e**) X-ray diffraction pattern of UiO-66 monolithic spheres. (**f**) Nitrogen physisorption isotherm (77 K) of monolithic UiO-66 spheres (full symbols = adsorption branch; open symbols = desorption branch).

**Table 1 | Porosity data for UiO-66 gels and powder.** $a_{BET}$ = BET surface area; $a_{micro}$ = micropore surface area, calculated using the t-plot method; $a_{ext}$ = external surface area ($a_{ext} = a_{BET} - a_{micro}$); $V_{pore}$ = total pore volume; $V_{micro}$ = micropore volume, calculated using the t-plot method.

| UiO-66 | $a_{BET}$ $m^2 \cdot g^{-1}$ | $a_{micro}$ $m^2 \cdot g^{-1}$ | $a_{ext}$ $m^2 \cdot g^{-1}$ | $V_{pore}$ $cm^3 \cdot g^{-1}$ | $V_{micro}$ $cm^3 \cdot g^{-1}$ |
|---|---|---|---|---|---|
| powder | 1167 | 1085 | 82 | 0.47 | 0.40 |
| xerogel | 1459 | 844 | 615 | 2.09 | 0.34 |
| xerogel sphere | 1127 | 403 | 724 | 1.90 | 0.17 |
| aerogel | 1255 | 847 | 408 | 1.66 | 0.33 |
| 10 nm nanoparticles[60] | 1181 | 730 | 451 | / | 0.29 |

# Supplementary Information

# Gel-Based Morphological Design of Zirconium Metal-organic Frameworks

Bart Bueken,[1] Niels Van Velthoven,[1] Tom Willhammar,[2,3] Timothée Stassin,[1] Ivo Stassen,[1] David A. Keen,[4] Rob Ameloot,[1] Sara Bals,[2] Dirk De Vos,[1,*] and Thomas D. Bennett.[5,*]

[1] Centre for Surface Chemistry and Catalysis, Department M²S, KU Leuven, Celestijnenlaan 200F p.o. box 2461, 3001 Heverlee, Belgium

[2] EMAT, University of Antwerp, Groenenborgerlaan 171, 2020 Antwerp, Belgium

[3] Department of Materials and Environmental Chemistry, Berzelii Center EXSELENT on Porous Materials, Stockholm University, S-106 91 Stockholm, Sweden

[4] ISIS Facility, Rutherford Appleton Laboratory, Harwell Campus, Didcot, Oxon OX11 0QX, United Kingdom

[5] Department of Materials Science and Metallurgy, University of Cambridge, 27 Charles Babbage Road, Cambridge CB3 0FS, United Kingdom

## Table of Contents

# Oil-drop granulation setup for monolithic UiO-66 spheres

A schematic overview of the in-house constructed oil-drop granulation setup is provided in Figure S0. The setup consists of a perfusor pump (Braun; Figure S0, a) which dispenses gel droplets at a fixed flow rate of 3 mL·h$^{-1}$ from a 3 mL syringe (BD Luer Lock) into a glass hopper (Figure S0, b and right panel) containing an immiscible silicone oil (Roger Coulon M1028/50). The end of the dispensing needle was blunted to form perfectly spherical gel droplets. A gear pump (Modelcraft #F3007; Figure S0, e) pumps the silicone oil throughout the setup at a constant flow rate, transporting the gel droplets through a 16 m polytetrafluoroethylene (PTFE) tube, which is submerged in a hot oil bath at 150 °C (Figure S0, c). The average residence time of the gel droplets in the hot PTFE tube is approximately 10 min, during which the droplets undergo a temperature-induced syneresis. Subsequently, the hardened monolithic spheres exit the PTFE tube and are captured in a Schott bottle (Figure S0, d) which has been modified to allow easy recovery of the spheres from the bottom of the vessel. The residual silicone oil is then recirculated by the gear pump to the glass hopper.

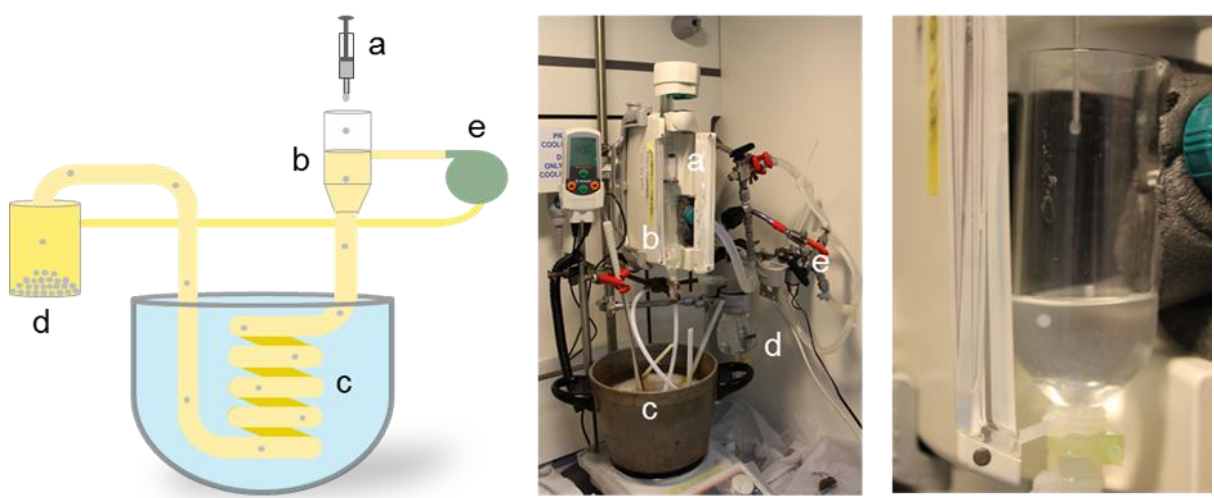

**Figure S0 | Oil-drop granulation setup.** Left & Middle: Schematic representation (left) and image (middle) of the setup used to prepare monolithic UiO-66 spheres. (a) perfusor pump; (b) hopper (b); (c) PTFE tube in oil bath; (d) Recovery of spheres in modified Schott bottle; (e) gear pump (e). Right: Close-up of the glass hopper (b) and dispensing needle (a), showing one gel droplet forming on the needle and one gel droplet suspended in the hopper.

## Additional images of UiO-66 monoliths

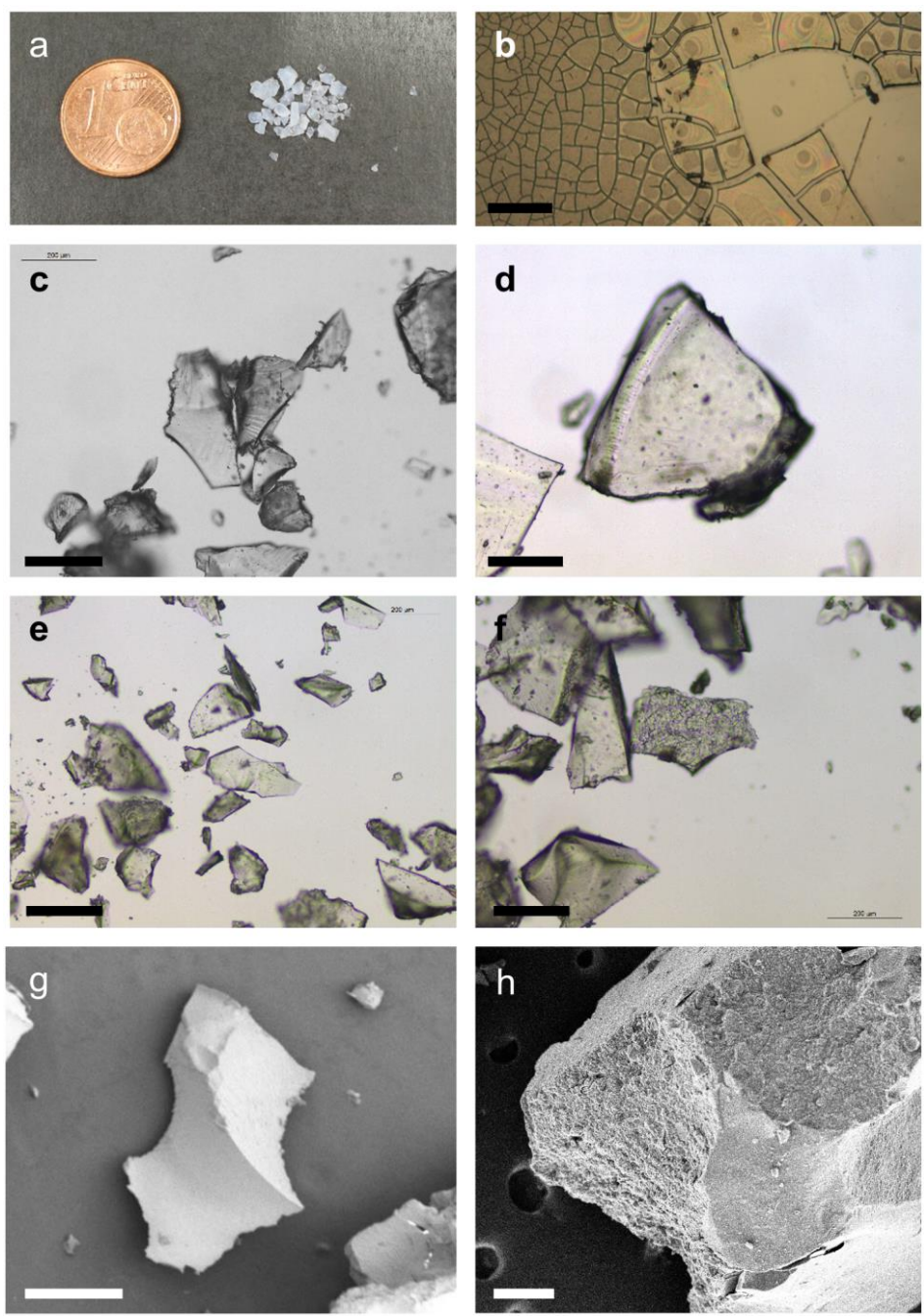

**Figure S1 | UiO-66 monolithic xerogels.** (**a**) millimetre-sized xerogel particles obtained after drying. (**b**) Transparent, mudcracked film obtained after drying a gel film coated on a glass petri dish (scale bar = 100 μm). (**c-f**) Optical images of transparent xerogel particles obtained after crushing particles as in (a) (scale bars = 200 μm). (**g-h**) Scanning electron micrographs of xerogel particles (scale bars = 50 μm (g); 200 μm (h)).

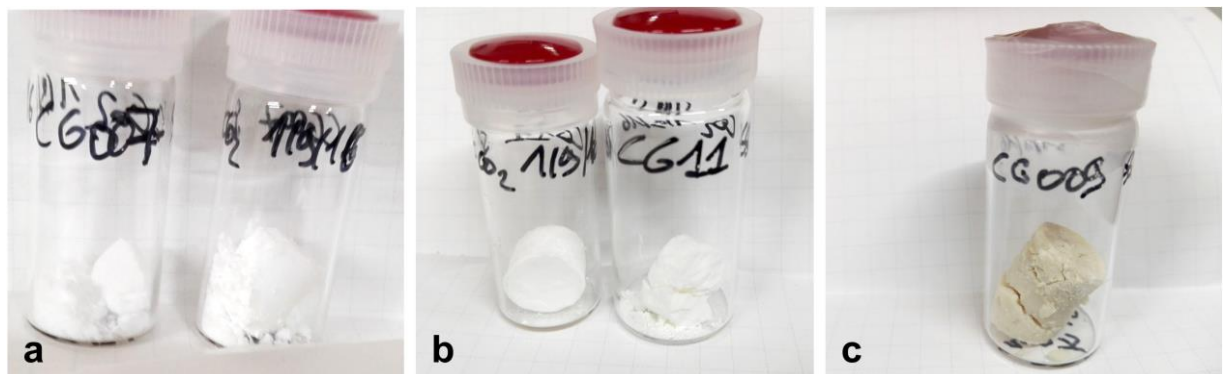

**Figure S2 | Various monolithic aerogels.** (**a**) UiO-66. (**b**) UiO-67. (**c**) UiO-66-NH$_2$.

# Pair Distribution Functions

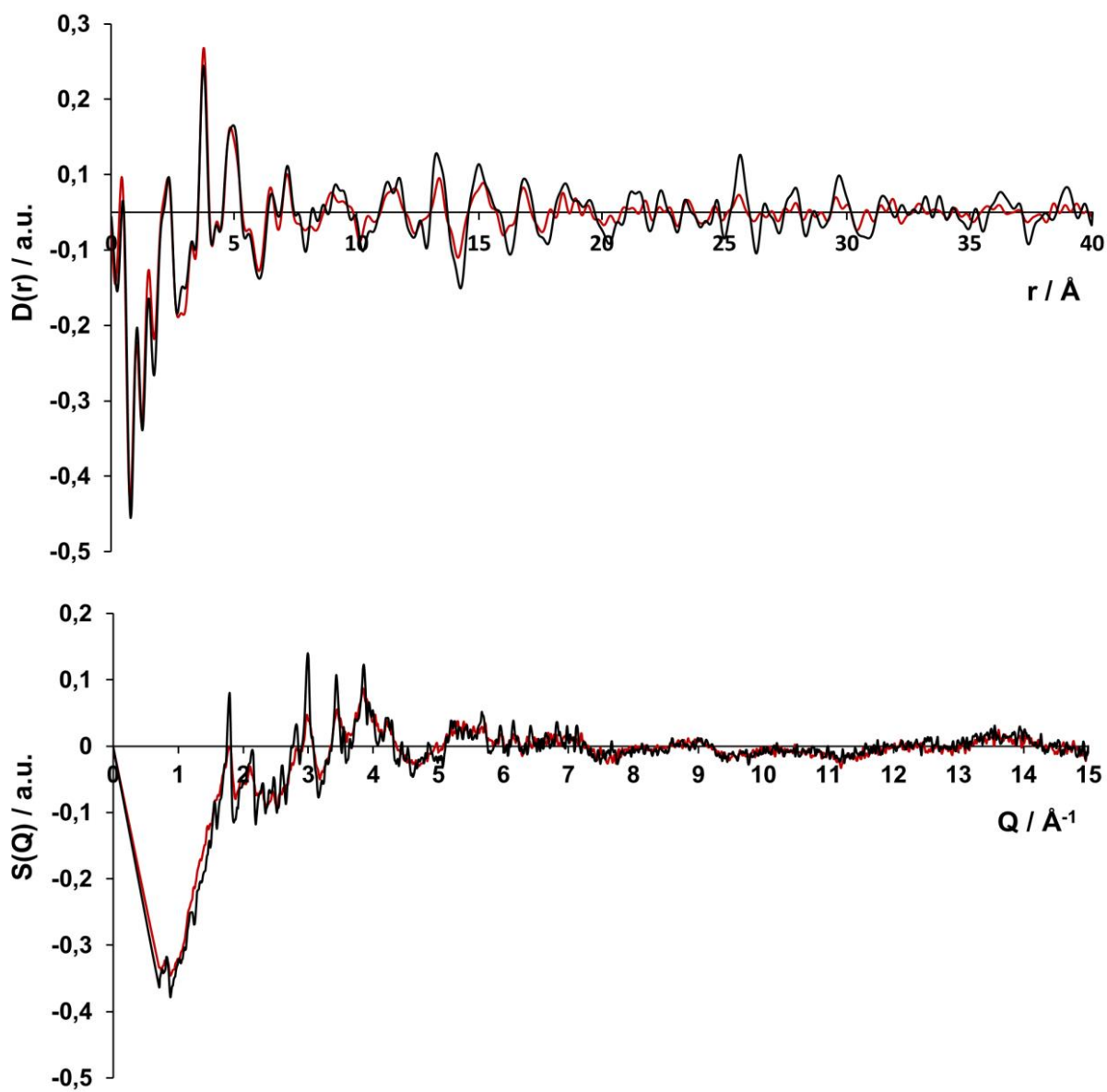

**Figure S3 | Pair distribution functions.** Top: Pair distribution functions (PDFs; D(r) vs. r) for a microcrystalline UiO-66 sample (black) and a monolithic xerogel of UiO-66 (red; Table S1, entry 10). The main peaks, at interatomic distances of 2.33 Å, 3.75 Å and 4.89 Å, correspond to UiO-66's intracluster Zr-O and Zr-Zr atom pairs, while those at greater interatomic distances can be attributed to correlations between atoms in neighbouring clusters. Bottom: Recorded structure factors S(Q) used to generate the PDFs by Fourier transformation for the microcrystalline UiO-66 sample (black) and the UiO-66 xerogel (red).

## Pore size distributions

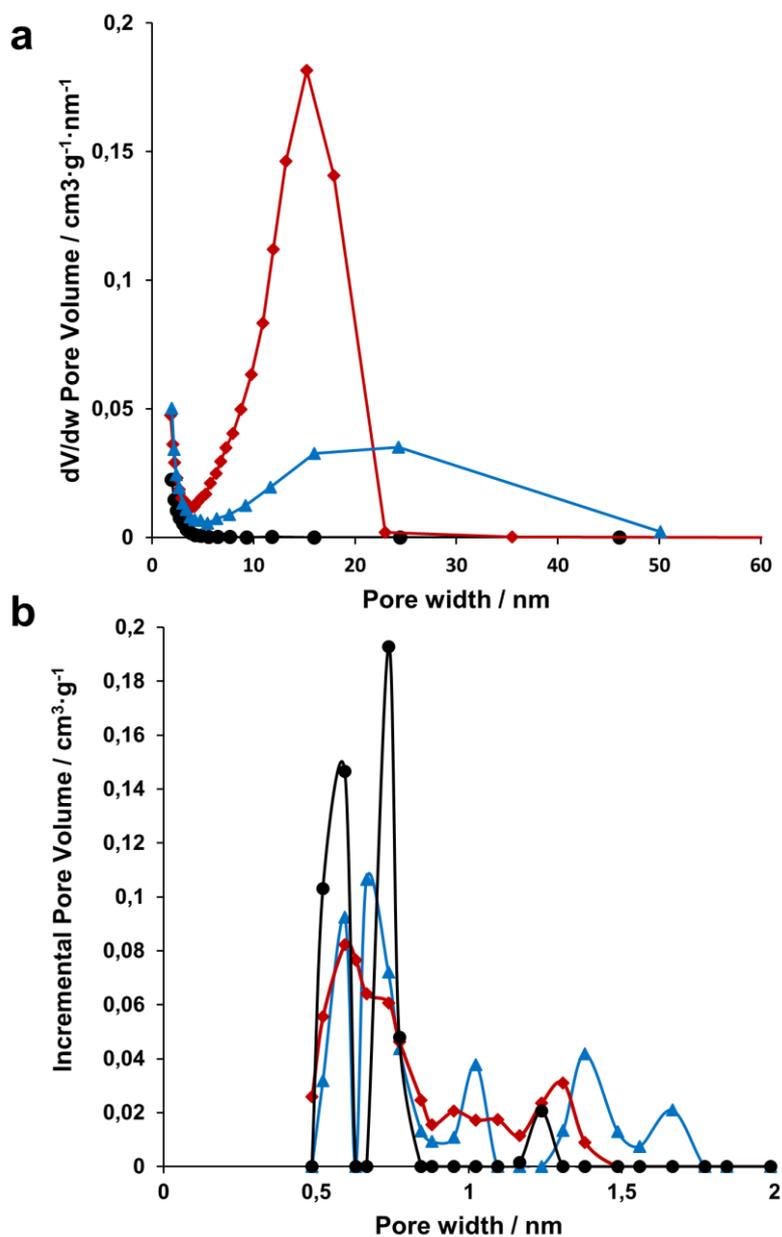

**Figure S4 | Pore size distribution.** (**a**) Mesopore size distribution obtained from the Barrett-Joyner-Halenda (BJH) model applied to the $N_2$ physisorption isotherms displayed in Figure 4, c. Hierarchical mesopores can be found in the UiO-66 xerogel (red diamonds) and aerogel (blue triangles). Note that in microcrystalline UiO-66 (black circles), no mesopores are observed. (**b**) Micropore size distribution obtained from the Tarazona Non-Local Density Functional Theory model applied to the $N_2$ physisorption isotherms displayed in Figure 4, c, illustrating the presence of micropores characteristic for UiO-66, with pore widths of 6 Å and 8 Å.[1]

## Thermogravimetric analyses

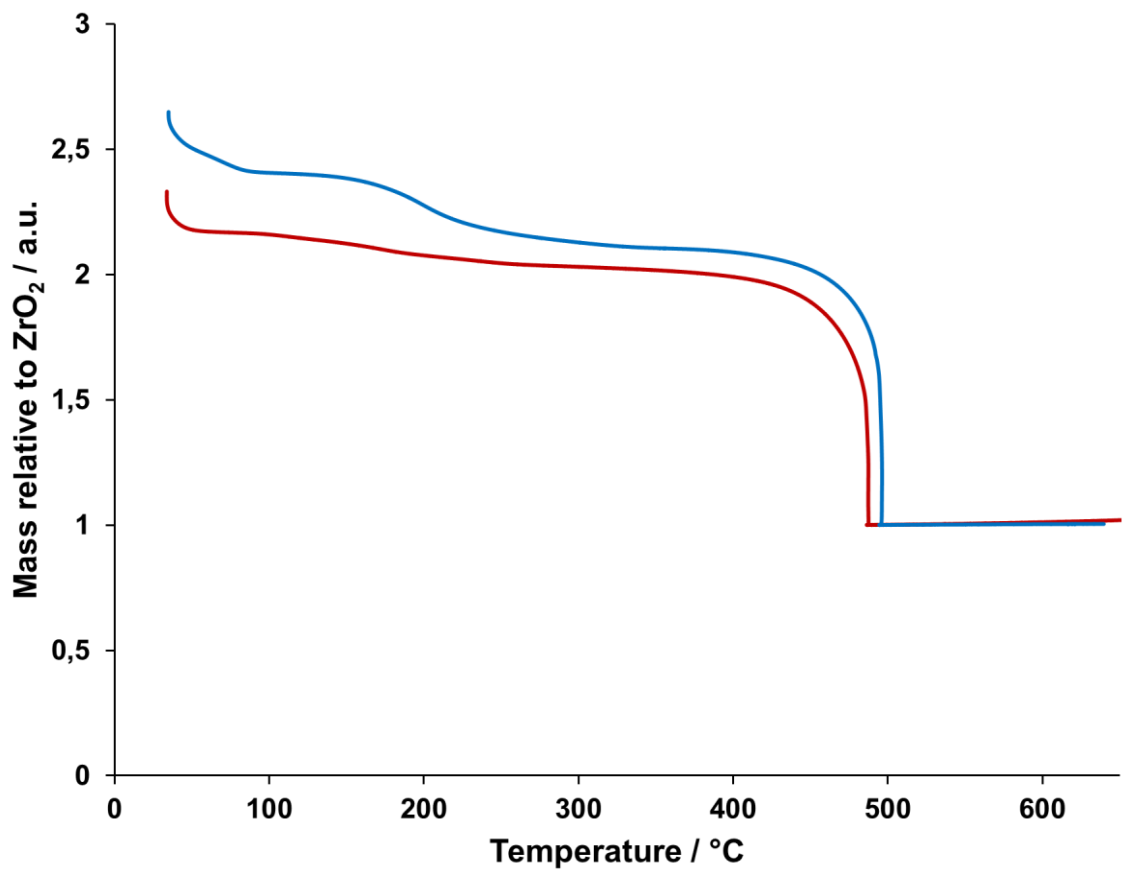

**Figure S5 | Thermogravimetric plots for UiO-66 gels.** Thermogravimetric plots for UiO-66 xerogel (red) and aerogel (blue) (both prepared from Table S1, entry 10). For the xerogel, an average of 11 linkers per $Zr_6$-cluster is found, while the aerogel has an average of 11.8 linkers per cluster, possibly due to unremoved $H_2$bdc.

# Nanoindentation experiments

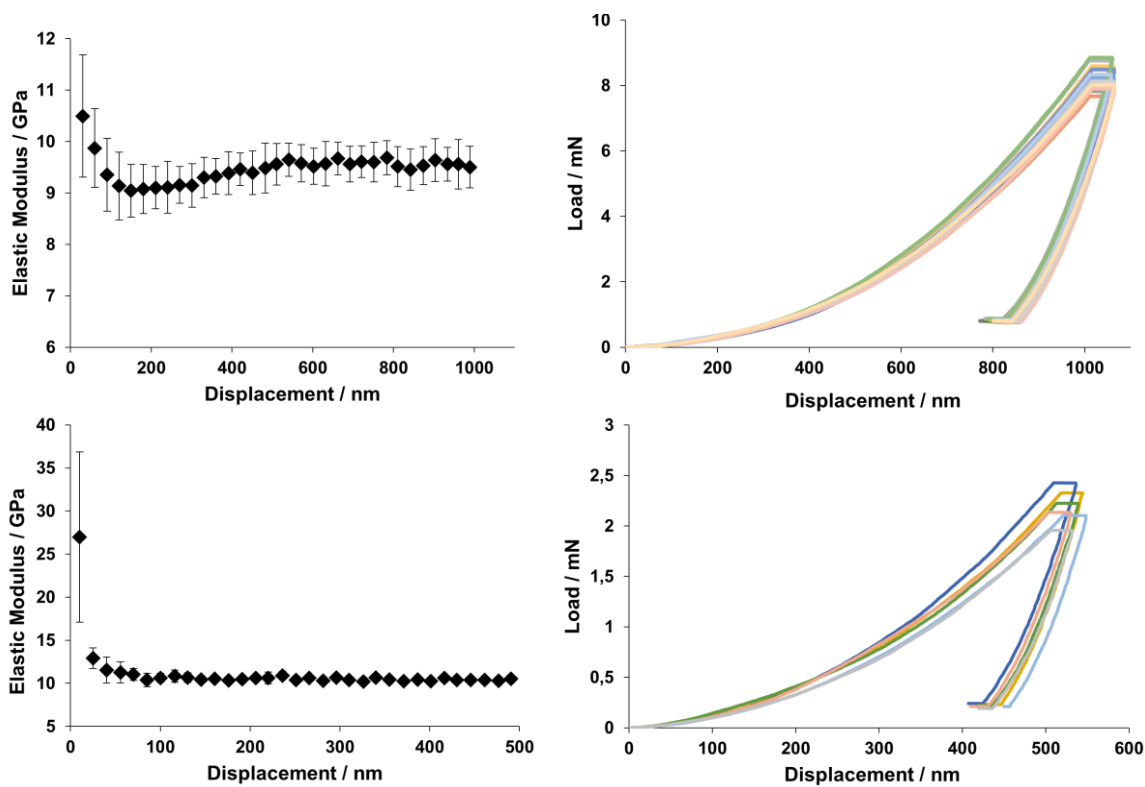

**Figure S6 | Nanoindentation experiments on monolithic UiO-66 xerogels.** Nanoindentation experiments were conducted on two xerogel samples (top curves = Sample 1; bottom curves = Sample 2; samples prepared from Table S1, entry 10). Left: Average elastic modulus (*E*) as a function of tip displacement into the monolith surface. Sample 1 showed an average *E* of 9.3 (±0.3) GPa, while for Sample 2 *E* averaged 10.5 (±0.5) GPa. Right: Load-displacement curves for the corresponding indentation experiments.

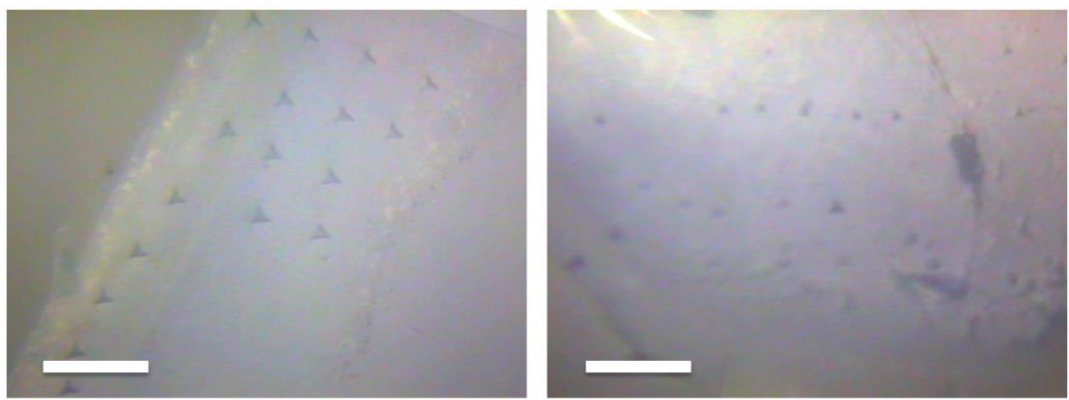

**Figure S7 | Optical microscopy of monolith nanoindentation.** Optical microscopy images of xerogel particles (left: Sample 1 from Figure S6; right: Sample 2 from Figure S6) after nanoindentation. Each triangular mark corresponds to single load-displacement experiment using a Berkovich tip (scale bar = 50 μm).

# Overview of gel formation

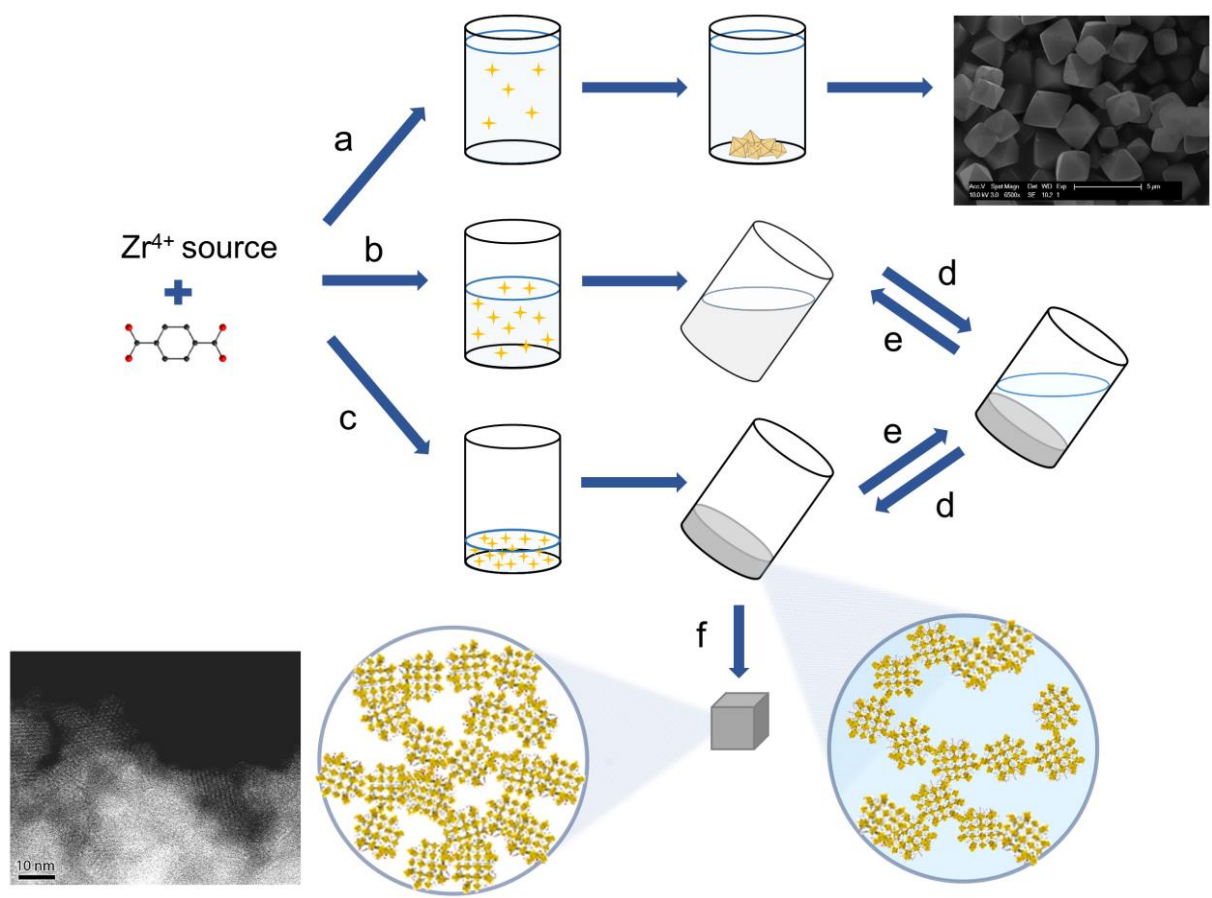

**Figure S8 | schematic overview of gel formation.** Synthesis of UiO-66 in dilute conditions, with a limited amount of water (a) leads to the nucleation and growth of microcrystalline particles. Conversely, high concentrations of linker and $Zr^{4+}$ source, in the presence of copious amounts of water (b & c) stimulates formation of the $[Zr_6O_4(OH)_4]^{12+}$ clusters and nucleation of the Zr-MOF (yellow stars). This leads to the formation of gel-like viscous colloidal suspensions of Zr-MOF nanoparticles, in which further crystal growth is hampered. At intermediate concentrations (b), 'flowing' gels can be observed, in which the number of interparticle interactions is insufficient to prevent gravitational flow. On the other hand, high nanoparticle concentrations (c) lead to viscoelastic 'non-flowing' gels with a network of weakly aggregated nanoparticles throughout the entire solvent volume. Tuning the nanoparticle concentration can interconvert the system between these two states. This can be achieved for instance by a centrifugation-driven syneresis of the 'flowing' gel (d) (e), or by diluting the 'non-flowing' gel through a shear-induced dispersion in additional solvent. Solvent removal from the 'non-flowing' gels yields monolithic xerogels or aerogels, depending on the drying conditions, consisting of randomly packed nanoparticles with interparticle mesopores.

## Gelation of various Zr-MOFs

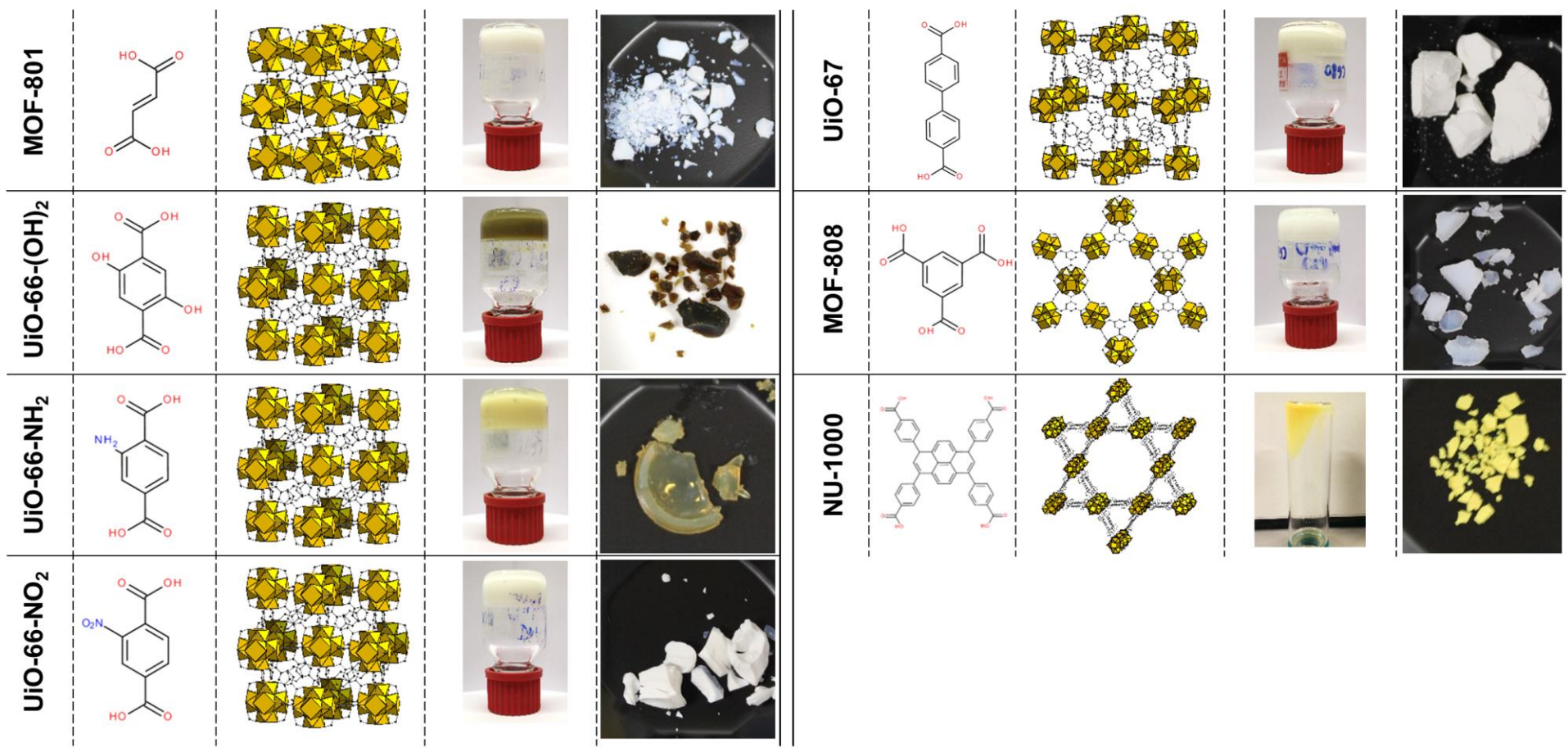

**Figure S9 | Gelation of various Zr-MOFs.** Gels and xerogel monoliths of various Zr-MOFs based on $Zr_6$-clusters.

# X-ray diffraction patterns of various Zr-MOF gels

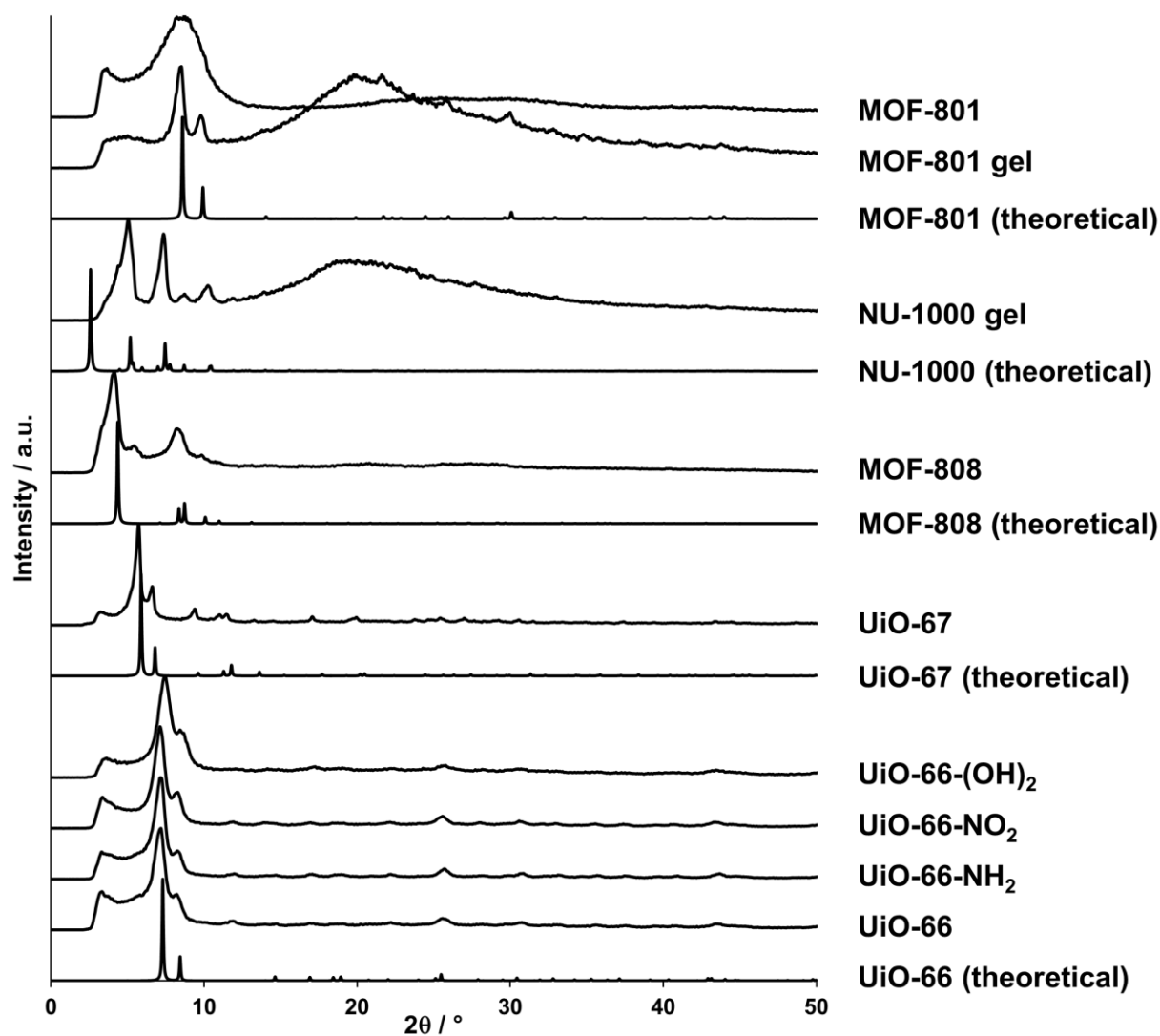

**Figure S10 | X-ray diffractograms of Zr-MOF (xero)gels.** X-ray diffraction patterns of monolithic xerogels of functionalized UiO-66 analogues, UiO-67 and MOF-808. For NU-1000, the diffraction pattern represents that of an ethanol-exchanged gel, due to the insufficient sample amount produced in the drying stage, which precluded recording a diffraction pattern of the monolithic material. The diffractometer setup features a primary beam stop up to ~2.8 ° 2θ, masking NU-1000's first reflection at 2.6 ° 2θ. For MOF-801 (Zr-fumarate), the diffraction pattern of the xerogel shows a loss of crystallinity relative to that of the ethanol-exchanged gel.

# Overview of UiO-66 gel syntheses

**Table S1 | Overview of UiO-66 gel syntheses.** Influence of Zr source, reactant concentration (DMF:Zr ratio) and additives on the macroscopic outcome of UiO-66 syntheses (2 h, 100 °C). $H_2O_{tot}$ = molar ratio of $H_2O$ in $ZrOCl_2 \cdot 8H_2O$ and additionally supplied, relative to Zr; $H_2O$ = molar ratio of added $H_2O$ (pure and as HCl 37 wt% solution) to Zr; $Cl_{tot}$ = molar ratio of $Cl^-$ in Zr source and additionally added relative to Zr; HCl = molar ratio of added HCl to Zr; AA = molar ratio of added acetic acid to Zr; MP indicates the formation of a microcrystalline powder; FG describes a 'flowing' gel-like, translucent to opaque suspension of moderate to high viscosity; NFG refers to a 'non-flowing', opaque gel for which the synthesis vessel can be turned upside down without the gel flowing downwards, see Figure 1, a). * = synthesized at 120 °C.

| #   | Outcome | Zr source | DMF:Zr | $H_2O_{tot}$ | $H_2O$ | $Cl_{tot}$ | HCl | AA |
|-----|---------|-----------|--------|--------------|--------|------------|-----|-----|
| 1   | MP      |           | 1503   | 14.8         | 6.8    | 4          | 2.0 | 0.0 |
| 2   | MP      |           | 620    | 14.2         | 6.2    | 3.8        | 1.8 | 3.5 |
| 3   | FG      |           | 620    | 42.7         | 34.7   | 3.8        | 1.8 | 3.5 |
| 4   | FG      |           | 388    | 14.3         | 6.3    | 3.8        | 1.8 | 0.0 |
| 5   | FG      |           | 388    | 14.3         | 6.3    | 3.8        | 1.8 | 3.5 |
| 6   | FG      |           | 388    | 42.9         | 34.9   | 3.8        | 1.8 | 3.5 |
| 7   | FG      | $ZrOCl_2 \cdot 8H_2O$ | 201 | 14.9 | 6.9 | 4 | 2.0 | 3.5 |
| 8   | NFG     |           | 155    | 14.2         | 6.2    | 3.8        | 1.8 | 3.5 |
| 9   | NFG     |           | 78     | 14.3         | 6.3    | 3.8        | 1.8 | 0.0 |
| 10  | NFG     |           | 78     | 14.3         | 6.3    | 3.8        | 1.8 | 3.5 |
| 11* | NFG     |           | 78     | 14.3         | 6.3    | 3.8        | 1.8 | 3.5 |
| 12  | NFG     |           | 78     | 37.4         | 29.4   | 2          | 0.0 | 3.5 |
| 13  | NFG     |           | 78     | 8.0          | 0      | 2          | 0.0 | 3.5 |
| 14  | NFG     |           | 65     | 34.0         | 26.0   | 9.5        | 7.5 | 0.0 |
| 15  | NFG     |           | 39     | 14.2         | 6.2    | 3.8        | 1.8 | 3.5 |
| 16  | MP      |           | 1501   | 1.0          | 1.0    | 4.0        | 0.0 | 0.0 |
| 17  | MP      |           | 621    | 14.3         | 14.3   | 4.0        | 0.0 | 3.5 |
| 18  | MP      |           | 388    | 6.2          | 6.2    | 5.8        | 1.8 | 0.0 |
| 19  | MP      |           | 388    | 6.3          | 6.3    | 5.8        | 1.8 | 3.5 |
| 20  | MP      |           | 388    | 14.3         | 14.3   | 5.8        | 1.8 | 3.5 |
| 21  | MP      |           | 388    | 40.0         | 40.0   | 5.8        | 1.8 | 3.5 |
| 22  | FG      |           | 388    | 120.5        | 120.5  | 5.8        | 1.8 | 3.5 |
| 23  | FG      | $ZrCl_4$  | 388    | 241.1        | 241.1  | 5.8        | 1.8 | 3.5 |
| 24  | MP      |           | 201    | 6.3          | 6.3    | 4.0        | 0.0 | 3.5 |
| 25  | NFG     |           | 201    | 43.2         | 43.2   | 4.0        | 0.0 | 3.5 |
| 26  | MP      |           | 155    | 14.2         | 14.2   | 4.0        | 0.0 | 3.5 |
| 27  | NFG     |           | 86     | 37.0         | 37.0   | 4.0        | 0.0 | 3.5 |
| 28  | /       |           | 78     | 0.0          | 0.0    | 4.0        | 0.0 | 3.5 |
| 29  | MP      |           | 78     | 7.6          | 7.6    | 6.0        | 2.0 | 0.0 |
| 30  | FG      |           | 78     | 19.0         | 19.0   | 4.0        | 0.0 | 3.5 |
| 31  | NFG     |           | 78     | 57.0         | 57.0   | 4.0        | 0.0 | 3.5 |